\documentclass[12pt]{article} 

\usepackage[english]{babel}

\usepackage{times}

\usepackage[labelfont=bf,figurename=Fig.,labelsep=period,font=footnotesize]{caption}

\usepackage{amsmath}
\usepackage{graphicx}
\usepackage[colorlinks=true, allcolors=blue]{hyperref}

\usepackage{comment}

\date{}

\usepackage{xcolor}
\usepackage{soul}
\definecolor{reddish}{HTML}{FBB4AE}
\definecolor{blueish}{HTML}{B3CDE3}
\definecolor{magentish}{HTML}{FF00AA}
\definecolor{greenish}{HTML}{a1d99b}

\title{\Large\bf Minorities in networks and algorithms}

\author
{
Fariba Karimi${}^{1*}$, Marcos Oliveira${}^{2}$, Markus Strohmaier${}^{4,3,1}$ \vspace{.1in}\\
\footnotesize{${}^{1}$Complexity Science Hub Vienna, Vienna, Austria}\\
\footnotesize{${}^{2}$Computer Science, University of Exeter, Exeter, United Kingdom}\\
\footnotesize{${}^{3}$Computational Social Science, GESIS--Leibniz Institute for the Social Sciences, Cologne, Germany}\\
\footnotesize{${}^{4}$Business School, University of Mannheim, Mannheim, Germany}\\
\footnotesize{$^\ast$Corresponding author; E-mail: karimi@csh.ac.at}
}

\begin{document}
\maketitle

\begin{abstract}


In this chapter, we provide an overview of recent advances in data-driven and theory-informed complex models of social networks and their potential in understanding societal inequalities and marginalization. We focus on inequalities arising from networks and network-based algorithms and how they affect minorities. In particular, we examine how homophily and mixing biases shape large and small social networks, influence perception of minorities, and affect collaboration patterns. We also discuss dynamical processes on and of networks and the formation of norms and health inequalities. Additionally, we argue that network modeling is paramount for unveiling the effect of ranking and social recommendation algorithms on the visibility of minorities. Finally, we highlight the key challenges and future opportunities in this emerging research topic.

\end{abstract}

\section*{Introduction}
Social inequalities---structured and recurrent patterns of unequal distribution of wealth, opportunities, and rewards for different social positions or statuses within a group or society---are on the rise, and current solutions to these problems are insufficient or failing~\cite{butler2006understanding}. For example, during the COVID-19 pandemic, we witness growing inequality in the mortality rate of minorities and marginalized groups, which is due to existing structural inequalities~\cite{bambra2020covid}. \textit{Structural Inequality} is one of the complex manifestations of social inequalities in which institutions, policies, and societies create privilege systems resulting from a confluence of unequal relations in roles, functions, decisions, rights, and opportunities~\cite{stolte1977structural}, which ultimately produces \textit{structural barriers} to equality and inclusiveness. 





\paragraph*{Minorities, marginalization, and inequalities.}

Before discussing this chapter's central topic, it is essential to define some of the fundamental sociological concepts we use throughout the text. Here, we use the notion of \textit{minority} to refer to a group of people who share a similar attribute (e.g., gender, race, ethnicity) and whose size is smaller than other groups' size. This imbalance in population size often results in inequalities and marginalization. \textit{Marginalization} is defined as ``to relegate to an unimportant or powerless position within a society or group". \textit{Social inequalities} refer to structured and recurrent patterns of unequal distribution of wealth, opportunities, and rewards for different social positions or statuses within a group or society~\cite{butler2006understanding}. In political philosophy and economics, scholars distinguish two types of equality: \textit{equality of opportunity} and \textit{equality of outcome}~\cite{fleurbaey1995equal}.
While we don't aim to discuss these concepts in depth, it is worth noting that in many cases, the inequalities that emerge from networks affect directly the former. 

\paragraph*{Emergence of inequalities and biases as a complex interacting system.}

Social networks are complex, time-dependent, and path-dependent systems comprising heterogeneous individuals interacting in evolving adaptive multi-layer networks of friendship, collaboration, communications, and trade, among other social interactions~\cite{walby2007complexity,holovatch2017complex}. Network models, despite being often stylized, have provided valuable insights into the emergence of macro-level phenomena such as riots~\cite{watts2002simple}, political movements~\cite{galam2002minority}, diffusion of innovations~\cite{young2011dynamics}, and predicting and preventing disease outbreaks~\cite{della2020network,haug2020ranking}. Social mechanisms can be inserted in network models (i.e., mechanistic models) in order to test hypothetical scenarios, evaluate the consistency of descriptive theories, explore emergent phenomena, and predict outcomes~\cite{holme2015mechanistic,hedstrom2010causal}. 

Much of the developments in computational models of complex networks have primarily focused on universality, phase transitions, and the emergence of scaling~\cite{holovatch2017complex}. Such a focus has often overlooked that social systems comprise individuals and groups with fundamentally heterogeneous attributes and historical precedent. These heterogeneities at the individual level and path dependencies can significantly impact how people interact and organize their social relationships. 

Social distinction and social differentiation can cause stratification in a complex society, where gender, race, and ethnicity are among the most pervasive characteristics (or attributes) that people use to make distinctions in terms of roles and expectations~\cite{grusky2019social,jost2015world}. Additionally, individual-level intentions and cognitive biases toward group members have been established and discussed in social psychology literature extensively~\cite{fletcher1990bias,brewer1979group}. However, the networks' impact on population-level bias and inequality remains relatively unexplored in the literature. Despite this research deficit, previous works have already established the plausibility of network effects on inequality by examining network influence, homophily, and diffusion processes, providing insights for computational modelling and empirical data-driven research~\cite{dimaggio2012network,garip2021network}. 

\paragraph*{Human biases and their manifestations in networks.}
There are various ways human biases manifest and exacerbate in social networks. Examples of those biases include social bias, population bias, self-selection bias, and historical bias~\cite{mehrabi2021survey}. Biases in networks can lead to unfairness in society and algorithms. For example, \textit{historical bias} can lead to the formation of an old boys club in scientific collaboration or for search engines to show predominantly male CEOs~\cite{suresh2019framework}. \textit{Social bias}, also known as attributional error~\cite{fletcher1990bias}, is also closely related to the network effect where other people's actions in our network proximity affect our judgment and choices. This phenomenon can directly affect whom we decide to create a link with or which information to share with others. \textit{Popularity bias} describes the human tendency to be attracted to popular items or individuals. For example, in search engines, the top-ranked results attract disproportionately more clicks/links than others, leading to cumulative advantages~\cite{baeza2018bias}. Many of these biases may have been driven by some biological and evolutionary processes that are essential for survival \cite{mcdermott2008evolutionary}. While we encourage the reader to reflect on those fundamental processes, any further elaboration is beyond the scope of this chapter.

\begin{figure}
    \centering
    \includegraphics[width=\columnwidth]{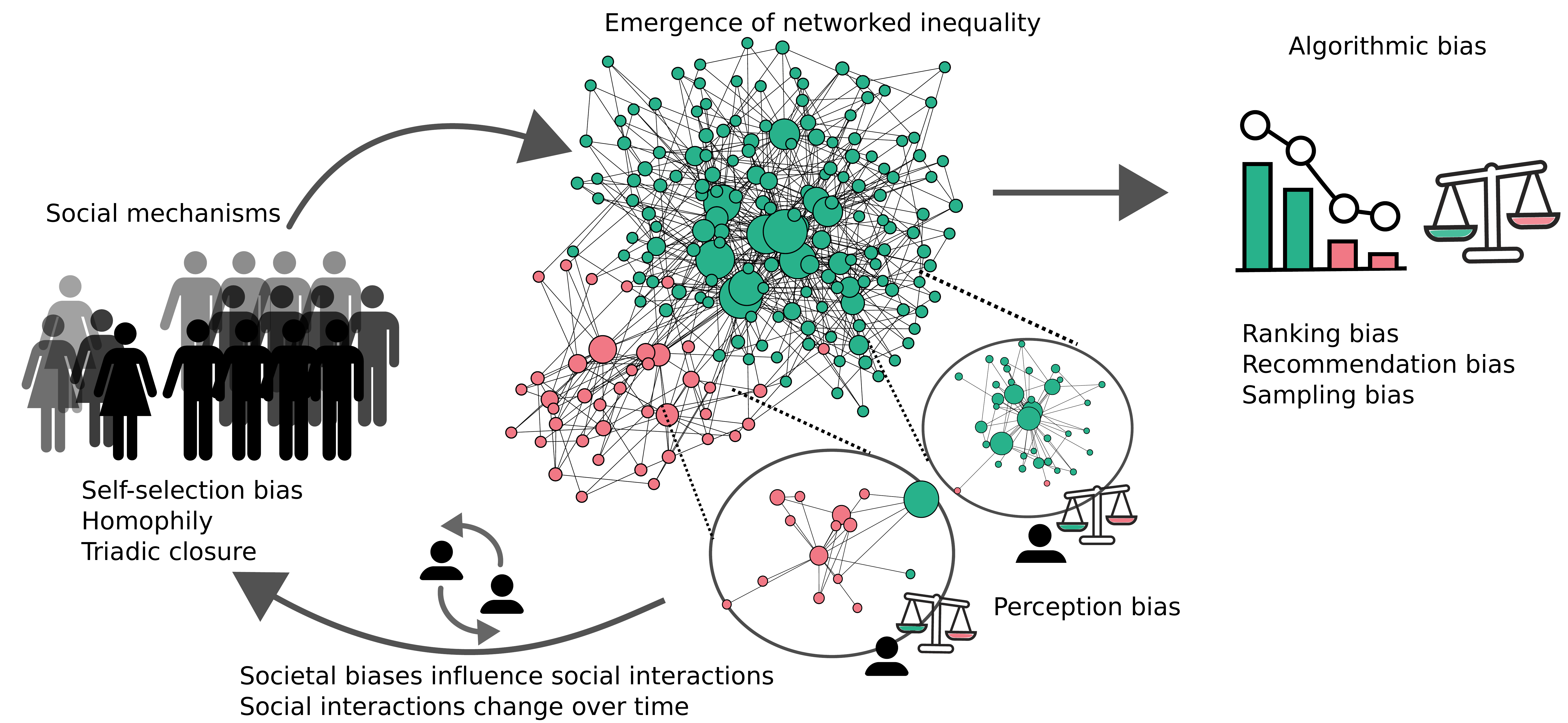}
    \caption{\textbf{Illustration of Network Inequality concept.} Social mechanisms in micro-level result in macro-level emergence of topological and structural conditions that affect dynamics on and of networks. The social networks are fed into algorithms and the inequality may be exacerbated by algorithms. The social networks are dynamic and change over time.}
    \label{fig:illustration}
\end{figure}

Figure \ref{fig:illustration} illustrates the \textit{Network Inequality} framework. Social mechanisms in micro-level results in macro-level emergence of topological and structural conditions that affect dynamical processes and network-based algorithms. 

We have organised the rest of this chapter in the following way. First, we discuss computational models to address societal issues such as visibility of minorities in large-scale social networks, minorities in small-scale face-to-face interactions, perception bias towards minorities, and minorities in collaborative environments. Second, we discuss visibility of minorities in network-based algorithms, the interplay between recommender systems and the network evolution, and in network sampling. Last, we discuss dynamics 'on' networks such as spreading of norms and diseases, and dynamics 'of' networks such as changes in social interactions over time and how these dynamics affect minorities. We close the paper by discussing key challenges and future directions.

\section*{Minorities in social networks}

Homophily---the idea that similarity breeds connection---is a fundamental mechanism of social tie formation~\cite{mcpherson2001birds}. Because individuals may identify themselves with groups based on categories such as gender, race, and ethnicity \cite{tajfel2004social}, homophily can drive the formation of mixing biases among those categories, resulting in groups of different sizes, which might affect individuals' visibility.


Visibility in networks is often defined as the attention given to an individual or a group compared to what we would expect from their prevalence or intrinsic value. Though this definition is somewhat limited (e.g., the assumption of group prevalence or intrinsic values is debatable), we can use mechanistic models to evaluate and experiment with such contested definitions systematically as a form of thought experiments (\textit{Gedankenexperiment}). 

Distorting the visibility of minorities via societal processes or algorithms can have severe consequences for society. This problem, known as \textit{invisibility syndrome}~\cite{franklin1999invisibility}, refers to situations in which minorities feel ignored and overlooked by the wider public. For example, social and academic invisibility prevents minorities from integrating into social networks and cultivating their potentials. The invisibility syndrome can have consequences for minorities. It can lead to minorities becoming silent, also known as the spiral of silence~\cite{noelle1974spiral}, leaving the community (e.g., dropping out from academia~\cite{wickware1997along,jadidi2017gender}), or in the extreme case, becoming violent towards the majority group because they do not have to conform to the same norm of the majority~\cite{bartlett2012edge}. 

\paragraph*{Mixing biases in large-scale networks.}
Now let us examine the visibility of minorities in large-scale social networks. We specifically focus on the effect of group imbalance and mixing biases (homophily) on the network visibility of minorities and majorities. To do so, we developed a simple preferential attachment network model with homophily (BA-Homophily model), including an analytical formulation~\cite{karimi2018homophily}. This model allows us to gain an in-depth insight into how nodes' degrees relate to group size and homophily. We use average group degree as a proxy for visibility, and find that homophily can penalize minorities by restricting their ability to connect to the bigger group. We found an unexpected tipping point, a non-sweet spot for minorities, in the mid-homophily range where they can be penalized the most. In the intermediate range of homophily, not only majority nodes prefer to attach to other majorities due to homophily, but also newly-arrived minority nodes prefer the majority because of their preferential attachment. As homophily increases, this dynamic changes, and minorities manage to recover their degree. In addition, by running a simple Susceptible-Informed diffusion dynamic, we observe that although strong homophilic tendencies help minorities recover their degree of connectivity, it will prevent them from accessing information from the majority group and vice versa. This insight allows us to quantitatively measure the advantages and disadvantages of mixing biases between the groups. Finally, we validated our model by analyzing different empirical networks and comparing them to the synthetic models.

\paragraph*{Mixing biases in small-scale networks.}
Beyond large-scale social networks that are useful to model virtual and cumulative interactions, in reality, we build meaningful and long-lasting connections via face-to-face interactions~\cite{SchaibleHandbook}. This essential human behavior is a primary form of social communication, inherited from our ancestors~\cite{dunbar_neocortex_1992}, and it is a fundamental mechanism for the transmission and affirmation of culture~\cite{Duncan1977}. 

Face-to-face interaction is particularly unique due to its spatial-temporal constraint: individuals must be at the same place and at the same time to interact. This constraint is critical to interaction opportunities, especially in confined situations, such as conferences and workplaces. For example, we only have a limited number of opportunities for interaction at conferences (e.g., coffee breaks). We found that this constraint, together with group mixing and size imbalance, can produce disparities in the visibility of social groups~\cite{oliveira2022group}. Once again, these disparities translate into members of specific social groups having fewer social ties than the average. We disentangle this phenomenon using a mechanistic model that embeds the properties of face-to-face interactions.

Due to the constraints in face-to-face interactions, when an individual uses an opportunity to interact with someone, fewer opportunities remain for interacting with other individuals. Depending on how groups mix, this limitation in opportunities leads to inequality. When a majority member interacts with someone from the majority, fewer opportunities remain for this individual to interact with minorities, decreasing minority connectivity. This inequality emerges irrespective of individuals' abilities. Even when minority members have a high intrinsic fitness (i.e., attractiveness), they can be at a disadvantage because group mixing affects how individuals identify other individuals' particularities. 

This systemic inequality is challenging to mitigate without the help of the largest group. Our model indicates that majority group mixing explains most of the variance in connectivity inequality. The minority group can only slightly reduce inequality, and a successful strategy for this reduction depends on group size. Our model predicts that, depending on the minority group size, minority members need to avoid connecting with the minority group to increase their connectivity. We found a tipping point that determines how minorities should strategize their social interactions. When the minority size is below a critical point, homophilic minority interaction decreases minority connectivity. Conversely, when the minority group size is sufficiently large, homophilic minority interaction helps in increasing minority connectivity. 


\paragraph*{Minorities and perception biases.}
The way people perceive the prevalence of others can influence their personal beliefs and behavior and shapes their collective dynamics. Many studies have documented biases in these social perceptions, including both overestimation and underestimation of the size of minority groups. We investigate whether such seemingly contradictory biases, namely false consensus (overestimation of the frequency of one’s attributes) and false uniqueness (underestimating the frequency of our view), can be explained merely by the structure of the social networks~\cite{lee2019homophily}. In other words, how does our estimate about minorities relates to our social networks? 

To address this question, first, we turned to examine this effect through a series of cross-cultural surveys. We asked questions about different societal attributes for which we knew the actual frequencies in the general population from existing national surveys in these countries. Participants answered three groups of questions. First, they answered questions about themselves (e.g., whether they smoke). Second, they estimated the frequency of people with each attribute in their networks (e.g., what fraction of your friends are smokers?). Third, we asked the participants to estimate the frequency of people with a specific attribute in the general population of their country. These questions allow us to estimate the homophily in personal networks and the perception of minorities. 

Next, we turned to our network model with homophily and examined whether we can explain the empirical observation in synthetic and real-world networks. Our results showed that when homophily is high, both minority and majority groups tend to overestimate their own size. In contrast, when homophily is low, people tend to overestimate the size of the minority group. More importantly, if we assume that people often aggregate their own perception with the perception of their neighbors (i.e., DeGroot weighted belief formalization), in homophilic networks, individuals cannot improve their perception because their peers do not add sufficient new information that would increase the accuracy of their estimates. However, in heterophilic networks, individuals benefit from considering their neighbors’ more diverse perceptions. 

\paragraph{Minorities and power.}
Though we often refer to minorities as under-represented groups who are disadvantaged due to the network processes, it is worth noting that some minorities, despite being smaller in size, disproportionately occupy powerful positions in networks. To understand how such scenarios happen, let us return to the scale-free homophilic networks that we described earlier in this chapter~\cite{karimi2018homophily}. In this model, mixing likelihood ranges from $0$ to $1$. The values below $0.5$ are heterophilic scenarios in which the probability of in-group ties is below 50\% resulting in minorities attracting the majorities. In such cases, because the majority group is larger and prefers to connect with the minority group, the minority members benefit from growing their number of connections---while the competition for receiving new connections due to the preferential attachment is limited due to their small size. The combination of heterophilic tendencies and preferential attachment results in inequality in the distribution of degree ranking in networks, as individuals in top ranks, disproportionately enjoy advantages as they stabilize their position in the rank and give few opportunities to others to catch up~\cite{ghoshal2011ranking}. This effect, minorities in central positions in the network is related to the \textit{majority illusion}~\cite{lerman_majority_2016} idea (a small powerful minority is percieved as the majority in the network). This is an important aspect, in particular, with respect to online ranking algorithms such as PageRank \cite{mariani2015ranking,espin2022inequality}. While the majority of the current works are focusing on the disadvantages that the algorithms create, it is crucial to reflect on the disparate gap that they can create for the individuals on the top of the ranks.

\paragraph*{Collaboration, success and dropouts of minorities.}
One important implication of network inequality is in academia. Science is largely a social process in which ideas spread between scholars, role models influence minds, and young scholars find their mentors and scientific interests through this process. In academia, collaboration patterns have a crucial impact on a scholar’s opportunities to establish an academic reputation and success \cite{sarig-a}. The diversity of collaboration \cite{alshebli2018preeminence}, the duration of collaboration \cite{petersen2015a} and the seniority and prominence of the collaborators \cite{petersen2014a} are among those patterns that can impact the academic career of young scholars. Structural and historical inequalities combined with the social process of imitation and adoption can create a condition in which minorities either do not enter certain areas of science or tend to have higher dropouts \cite{clifton2019mathematical}. Our analysis of the career age and collaboration patterns of more than one million computer scientists for more than four decades shows that women have higher dropouts throughout all career ages, and establish more links with other female researchers \cite{jadidi2017gender}, and since they are a minority, this phenomenon can potentially impact their visibility and ability to access novel ideas or information. 


\section*{Minorities and algorithms}


Artificial Intelligence and online algorithms have become cornerstones of many decision-making processes in industry, academia, and the public sector, from the diagnosis of diseases and the forecast of recidivism risk in criminal trials to the recommendation of whom to hire and whom to follow on social media \cite{kuziemski2020ai}. At the same time, these algorithms have recently come under public scrutiny because the relatively high levels of accuracy often come at the cost of biases, such as misclassifying under-represented groups or reinforcing pre-existing inequalities. For instance, they may contribute to systematic discrimination against women in the job market, or against Hispanic and Black people in justice systems \cite{dressel2018accuracy}. While the computer science community has tackled these issues by calibrating their models according to certain definitions of fairness \cite{zafar2017fairness,lepri2018fair}, understanding the influence of social network structure on algorithmic outcomes requires more attention and future research. 

Machine learning and AI systems often process existing input data in order to train their models for prediction tasks. Mechanistic network models from the domain of complexity science can enable researchers to consider various hypothetical scenarios that can occur in societies but are not available in input training data. This allows to evaluate the robustness of algorithms with regard to different aspects concerning minorities, for example fairness or discrimination. In the following, we illustrate a few examples in which network models are used to evaluate the fairness of algorithms in the presence of minorities. 


\paragraph*{Minorities and sampling.} When collecting and analysing social network data, online or offline, we are inherently conducting some form of sampling. For example, surveys or selecting a school to collect information about the students and their contacts are a form of random node sampling. In contrast collecting data based on phone calls is a form of random edge sampling. In social sciences, certain minorities are denoted as \textit{hard-to-reach} groups (e.g. homeless people or illegal immigrants) in which snowball sampling seems to be a promising method to access those populations. Using the network model with adjustable homophily and minority size described earlier, we evaluated different sampling techniques with respect to conserving the topological position of nodes (degree centrality) and the visibility of groups in top ranks \cite{wagner2017sampling}. Our results show that different network sampling techniques are highly sensitive with regard to capturing the expected centrality of nodes, and that their accuracy depends on relative group size differences and on the level of homophily that can be observed in the network. In other words, uninformed sampling from social networks with attributes can significantly impair
the ability of researchers to draw valid conclusions about the centrality of nodes in social networks. In a follow-up work, we examined how network sampling strategy and sample size affect accuracy of relational inference in networks \cite{espin2018towards}. Relational inference leverages relationships between entities and links in a network to infer information about the network from a small sample. We find that in heterophilic networks, we can obtain good accuracy when only small samples of the network are initially labeled, regardless of the sampling strategy. Surprisingly, this is not the case for homophilic
networks, and sampling strategies that work well in heterophilic networks lead to large inference errors. In summary this means that social network structures that are driven by homophily can induce biases on sampling and inference procedures.

\paragraph*{Minorities in network-based ranking and recommender systems.}

Many ranking and recommender systems in online algorithms are, based on, and sensitive to, the directionality of the links in the network. For example, PageRank centrality deployed by Google, or Who-to-Follow (WTF) link recommendation systems deployed in Twitter are both based on directed networks. In order to investigate the effect of directionality of the links and homophily on the ranking of minorities, we proposed a directed network model with preferential attachment and homophily and explored the influence of network structure on the rank distributions of these algorithms \cite{espin2022inequality}. Our findings, align with other similar works \cite{fabbri2020effect},  suggest that these two algorithms can reduce, replicate and amplify the representation of minorities in top ranks when majorities are homophilic, neutral and heterophilic, respectively. Moreover, when this representation is reduced, minorities may improve their visibility in the rank by connecting strategically in the network, by e.g., increasing their out-degree or homophily when majorities are also homophilic. Investigating such scenarios in directed networks are promising avenues that can inform how to design algorithms that achieve fairness. 

As a result of recommender algorithms, users may accept new social links, and hence, social network structures can change over time. Based on a series of synthetic network experiments, we showed that not only such algorithms tend to increase the clustering and cohesion within groups, they are also prone to favor nodes with high in-degree. This bias towards popular nodes may reinforce pre-existing social biases and contribute to cumulative advantages \cite{ferrara2022link}. 



\section*{Minorities and dynamics on and of networks}

Social interactions are time-dependent, dynamic, and they change as the social system evolves. These time dependencies can impact communities in networks~\cite{mucha2010community}, diffusion of information or norms, and causality direction~\cite{holme2015modern}. Non-Markovian properties of temporal networks constitute an essential dimension of complexity in time-varying complex systems, and they can lead to different types of diffusion outcomes~\cite{scholtes2014causality}. For example, organizational burstiness (coordinated burst in social interactions such as a coffee break at a conference) may speed up the complex contagion by activating many social interactions in a short duration of time~\cite{karimi2013threshold,karimi2013temporal}.  

When certain disadvantaged groups arrive in a social system with delay or at a slower rate, due to historical discrimination or societal inequalities (e.g., women in universities~\cite{kong2021first}, or African-Americans in the blue-collar workforce~\cite{telles1994industrialization}), the diffusion speed and hierarchy climbing for such groups is slower compared to more established groups. This late-arrival mechanism affects the age of the ego-network of minorities, the cumulative advantages, and their network position. By understanding this dynamic, we recently showed that the gender disparity in citation patterns of female-led papers and male-led papers is related mainly to the first-mover advantage of men and not a direct act of discrimination against women~\cite{kong2021first}. Nevertheless, it is worth noting that such effects may create implicit biases and discrimination similar to what sociologists found as \textit{racism without racists} effects~\cite{bonilla2006racism} in which structural racism can exist without anyone being racist.

Apart from the dynamics \textit{of} networks, one can also consider dynamical processes that unfold \textit{on} networks. Here, we discuss two types of dynamical processes: (1) diffusion of norms, (2) disease spreading, and how these dynamics affect minorities. Social scientists argue that diffusion of norms and culture falls under the category of complex contagion while disease spreading processes are often simple contagion. Complex contagion describes situations in which multiple social links of an individual must adopt a new norm or behaviour to convince this individual to adopt it. The rationale behind these processes is that humans, as social animals, follow conformity and social imitation when faced with limited access to information. Mark Granovetter first proposed a model of contagion to explain the emergence of norms and social uprising~\cite{granovetter1978threshold}. The core principle is that individuals have a threshold or stubbornness to adopt a new behaviour. An individual will adopt a new behaviour when a certain fraction of people (within this individual's observation horizon) also adopt the behaviour. In large societies, it is safe to assume that our observation horizon is somewhat limited to our direct contacts\footnote{Here, we ignore online social media exposure.}. Borrowing this idea, in one of the early works in contemporary network science, Watts showed that being embedded in a large-scale social network, even if the network is completely random, leads to situations where contagion can occur~\cite{watts2002simple}. This finding points to the fact that even without any intelligent decision-making process or social engineering, social contagion can occur depending on how we are wired to others~\cite{karimi2015tightly}. However, given the fact that groups mix differently (i.e., mixing biases), as we discussed in the earlier section, how would those biases in mixing affect the formation of norms and culture? In what societies can minorities propagate their norms, and in what societies do they need to adopt the majority's norms? 

\paragraph{Minorities and norms.} 
Following the same network models we described earlier, we examine the Granovetter threshold model on networks with the presence of the minorities and majorities and varying degrees of group mixing~\cite{kohne2020role}. We explore whether minorities can maintain their norms or even impose their norms on the majority group and in what situation normative conflict increases or decreases. Our first finding is that when norms align with the group categories (i.e., minorities disproportionately follow one norm while the majorities follow the opposing norm), friction and conflict occur while people need to swing in one way or another. Second, in societies with a moderate level of homophily, the majority norm takes over. Finally, as the minority group increases in size, it is more likely to retain its norm and influence the majority. Our simulation supports three strategies for maintaining minority cultural practices employed by minority groups in reality: isolationism, adopting positions of influence, and increasing the group size of one’s minority. In future work, it would be valuable to examine the influence of other mechanisms such as resistance, changes in the social interactions or heterogeneity in the group mixing.  

\paragraph{Minorities and health inequality.}  
Pandemics expose and deepen existing racial, ethnic, and income inequalities in society by aggravating resource constraints and rendering living conditions direr for those who live on the margins. During the recent COVID-19 pandemic, we witnessed a disproportionate infection rate among marginalized groups, people of color~\cite{bambra2020covid}, and low-income groups~\cite{adam2022pandemic}. Health inequalities are often due to existing structural inequalities in which societies perpetuate discrimination through mutually reinforcing systems of housing, education, employment, and health care~\cite{bailey2017structural}. By combining epidemic modelling in networks with agent-based simulation, our analysis has shown that within the lower-income population that cannot afford to quarantine themselves, the epidemic spreads faster and results in more death. In addition, since low-income individuals often live in segregated neighbourhoods that are densely populated, the epidemic has more chance of spreading from one person to another \cite{sajjadi2022structural}. These findings demonstrate that inequalities in one layer can diffuse to other layers of social lives, and it requires complexity thinking to understand and tackle them.

\section*{Key challenges and future directions}

In this chapter, we have reviewed recent data-driven, theory-informed research on inequality and minorities in social networks from the perspective of complexity science and computational social science. The research we presented helps us understand inequalities and marginalization in society; it sheds light on potential ways to alleviate such critical societal problems. More importantly, these research studies introduce novel challenges and opportunities. 

For example, network modelling and computational, agent-based simulations help us understand different dimensions of social interactions and structural inequalities. With this approach, we can disentangle existing inequalities, elaborate informed policies to improve the status quo, and prevent the formation of new inequalities (e.g., biased algorithms). However, we advocate that complex computational models of society should consider socio-demographic attributes of individuals, heterogeneity in group mixing, and how societies evolve and mix over time. In parallel to that, social theories of diffusion, tie formation mechanisms, and group identity should be modelled and validated with empirical data as much as possible~\cite{watts2017should}. 

Furthermore, we note that inequalities are multi-faceted and multi-dimensional. While a majority of the current research is on understanding the effect of gender or race on networks, evidence suggests that individuals who are in the intersection of multiple disadvantaged categories may be discriminated against and also experience inequalities more disproportionately compared to others~\cite{crenshaw1989demarginalizing}. Future research should pay more attention to this aspect.  

For data-driven research, we believe that mechanistic computational network models can play an instrumental role in order to validate competing theories, providing new theories and insights, and in particular, help machine learning methods when input data is missing or biased~\cite{steinbacher2021advances}. Understanding how biased data can affect computational methods is an essential future avenue for research.

Overall, we hope this chapter supplies readers with a broad overview of the diverse opportunities and challenges for understanding minority groups in social networks and serves as an inspiration for their own research agenda, aiming to transform society for the better.  

\section*{Acknowledgement}
We thank our key collaborators including Claudia Wagner, Eun Lee, Mirta Galesic, Lisette Espin-Noboa, Mathieu G{\'e}nois,  Mohsen Jadidi, Maria Zens, Antonio Ferrara, Julian Kohne, Natalie Gallagher, Sina Sajjadi, Samuel Martin-Gutierrez, Matthias Raddant, Hyunsik Kong, Petter Holme and Kristina Lerman, who helped in shaping many of these ideas and research papers. We thank Wayt Gibbs and Aleszu Bajak for their feedback on the manuscript. Fariba Karimi was supported by the Austrian research agency FFG, project number 873927. 

\bibliographystyle{naturemag}

\bibliography{references}

\end{document}